\documentclass[aps,prl,showpacs,twocolumn,groupedaddress]{revtex4}  

\usepackage{graphicx}  
\usepackage{dcolumn}   
\usepackage{bm}        
\usepackage{amssymb}   

\def\lsim{\mathrel{\rlap{\lower4pt\hbox{\hskip1pt$\sim$}}\raise1pt\hbox{$<$}}}
\def\gsim{\mathrel{\rlap{\lower4pt\hbox{\hskip1pt$\sim$}}\raise1pt\hbox{$>$}}}

\begin{document}


\hspace{5.2in} \mbox{FERMILAB-PUB-07-117-E}

\title{$Z\gamma$ production and limits on anomalous $ZZ\gamma$ and $Z\gamma\gamma$ couplings in $p\bar{p}$ collisions at $\sqrt{s}$~=~1.96~TeV}
%
\author{                                                                      
V.M.~Abazov,$^{35}$                                                           
B.~Abbott,$^{75}$                                                             
M.~Abolins,$^{65}$                                                            
B.S.~Acharya,$^{28}$                                                          
M.~Adams,$^{51}$                                                              
T.~Adams,$^{49}$                                                              
E.~Aguilo,$^{5}$                                                              
S.H.~Ahn,$^{30}$                                                              
M.~Ahsan,$^{59}$                                                              
G.D.~Alexeev,$^{35}$
% ...
\\                                                                            
\vskip 0.30cm                                                                 
\centerline{(D\O\ Collaboration)}                                             
\vskip 0.30cm                                                                 
}                                                                             
\affiliation{                                                                 
\centerline{$^{1}$Universidad de Buenos Aires, Buenos Aires, Argentina}       
\centerline{$^{2}$LAFEX, Centro Brasileiro de Pesquisas F{\'\i}sicas,         
                  Rio de Janeiro, Brazil}                                     
\centerline{$^{3}$Universidade do Estado do Rio de Janeiro,                   
                  Rio de Janeiro, Brazil}                                     
\centerline{$^{4}$Instituto de F\'{\i}sica Te\'orica, Universidade            
                  Estadual Paulista, S\~ao Paulo, Brazil}                     
\centerline{$^{5}$University of Alberta, Edmonton, Alberta, Canada,           
                  Simon Fraser University, Burnaby, British Columbia, Canada,}
\centerline{York University, Toronto, Ontario, Canada, and                    
                  McGill University, Montreal, Quebec, Canada}                
\centerline{$^{6}$University of Science and Technology of China, Hefei,       
                  People's Republic of China}                                 
\centerline{$^{7}$Universidad de los Andes, Bogot\'{a}, Colombia}             
\centerline{$^{8}$Center for Particle Physics, Charles University,            
                  Prague, Czech Republic}                                     
\centerline{$^{9}$Czech Technical University, Prague, Czech Republic}         
\centerline{$^{10}$Center for Particle Physics, Institute of Physics,         
                   Academy of Sciences of the Czech Republic,                 
                   Prague, Czech Republic}                                    
\centerline{$^{11}$Universidad San Francisco de Quito, Quito, Ecuador}        
\centerline{$^{12}$Laboratoire de Physique Corpusculaire, IN2P3-CNRS,         
                   Universit\'e Blaise Pascal, Clermont-Ferrand, France}      
\centerline{$^{13}$Laboratoire de Physique Subatomique et de Cosmologie,      
                   IN2P3-CNRS, Universite de Grenoble 1, Grenoble, France}    
\centerline{$^{14}$CPPM, IN2P3-CNRS, Universit\'e de la M\'editerran\'ee,     
                   Marseille, France}                                         
\centerline{$^{15}$Laboratoire de l'Acc\'el\'erateur Lin\'eaire,              
                   IN2P3-CNRS et Universit\'e Paris-Sud, Orsay, France}       
\centerline{$^{16}$LPNHE, IN2P3-CNRS, Universit\'es Paris VI and VII,         
                   Paris, France}                                             
\centerline{$^{17}$DAPNIA/Service de Physique des Particules, CEA, Saclay,    
                   France}                                                    
\centerline{$^{18}$IPHC, Universit\'e Louis Pasteur et Universit\'e           
                   de Haute Alsace, CNRS, IN2P3, Strasbourg, France}          
\centerline{$^{19}$IPNL, Universit\'e Lyon 1, CNRS/IN2P3, Villeurbanne, France
                   and Universit\'e de Lyon, Lyon, France}                    
\centerline{$^{20}$III. Physikalisches Institut A, RWTH Aachen,               
                   Aachen, Germany}                                           
\centerline{$^{21}$Physikalisches Institut, Universit{\"a}t Bonn,             
                   Bonn, Germany}                                             
\centerline{$^{22}$Physikalisches Institut, Universit{\"a}t Freiburg,         
                   Freiburg, Germany}                                         
\centerline{$^{23}$Institut f{\"u}r Physik, Universit{\"a}t Mainz,            
                   Mainz, Germany}                                            
\centerline{$^{24}$Ludwig-Maximilians-Universit{\"a}t M{\"u}nchen,            
                   M{\"u}nchen, Germany}                                      
\centerline{$^{25}$Fachbereich Physik, University of Wuppertal,               
                   Wuppertal, Germany}                                        
\centerline{$^{26}$Panjab University, Chandigarh, India}                      
\centerline{$^{27}$Delhi University, Delhi, India}                            
\centerline{$^{28}$Tata Institute of Fundamental Research, Mumbai, India}     
\centerline{$^{29}$University College Dublin, Dublin, Ireland}                
\centerline{$^{30}$Korea Detector Laboratory, Korea University,               
                   Seoul, Korea}                                              
\centerline{$^{31}$SungKyunKwan University, Suwon, Korea}                     
\centerline{$^{32}$CINVESTAV, Mexico City, Mexico}                            
\centerline{$^{33}$FOM-Institute NIKHEF and University of                     
                   Amsterdam/NIKHEF, Amsterdam, The Netherlands}              
\centerline{$^{34}$Radboud University Nijmegen/NIKHEF, Nijmegen, The          
                  Netherlands}                                                
\centerline{$^{35}$Joint Institute for Nuclear Research, Dubna, Russia}       
\centerline{$^{36}$Institute for Theoretical and Experimental Physics,        
                   Moscow, Russia}                                            
\centerline{$^{37}$Moscow State University, Moscow, Russia}                   
\centerline{$^{38}$Institute for High Energy Physics, Protvino, Russia}       
\centerline{$^{39}$Petersburg Nuclear Physics Institute,                      
                   St. Petersburg, Russia}                                    
\centerline{$^{40}$Lund University, Lund, Sweden, Royal Institute of          
                   Technology and Stockholm University, Stockholm,            
                   Sweden, and}                                               
\centerline{Uppsala University, Uppsala, Sweden}                              
\centerline{$^{41}$Physik Institut der Universit{\"a}t Z{\"u}rich,            
                   Z{\"u}rich, Switzerland}                                   
\centerline{$^{42}$Lancaster University, Lancaster, United Kingdom}           
\centerline{$^{43}$Imperial College, London, United Kingdom}                  
\centerline{$^{44}$University of Manchester, Manchester, United Kingdom}      
\centerline{$^{45}$University of Arizona, Tucson, Arizona 85721, USA}         
\centerline{$^{46}$Lawrence Berkeley National Laboratory and University of    
                   California, Berkeley, California 94720, USA}               
\centerline{$^{47}$California State University, Fresno, California 93740, USA}
\centerline{$^{48}$University of California, Riverside, California 92521, USA}
\centerline{$^{49}$Florida State University, Tallahassee, Florida 32306, USA} 
\centerline{$^{50}$Fermi National Accelerator Laboratory,                     
            Batavia, Illinois 60510, USA}                                     
\centerline{$^{51}$University of Illinois at Chicago,                         
            Chicago, Illinois 60607, USA}                                     
\centerline{$^{52}$Northern Illinois University, DeKalb, Illinois 60115, USA} 
\centerline{$^{53}$Northwestern University, Evanston, Illinois 60208, USA}    
\centerline{$^{54}$Indiana University, Bloomington, Indiana 47405, USA}       
\centerline{$^{55}$University of Notre Dame, Notre Dame, Indiana 46556, USA}  
\centerline{$^{56}$Purdue University Calumet, Hammond, Indiana 46323, USA}    
\centerline{$^{57}$Iowa State University, Ames, Iowa 50011, USA}              
\centerline{$^{58}$University of Kansas, Lawrence, Kansas 66045, USA}         
\centerline{$^{59}$Kansas State University, Manhattan, Kansas 66506, USA}     
\centerline{$^{60}$Louisiana Tech University, Ruston, Louisiana 71272, USA}   
\centerline{$^{61}$University of Maryland, College Park, Maryland 20742, USA} 
\centerline{$^{62}$Boston University, Boston, Massachusetts 02215, USA}       
\centerline{$^{63}$Northeastern University, Boston, Massachusetts 02115, USA} 
\centerline{$^{64}$University of Michigan, Ann Arbor, Michigan 48109, USA}    
\centerline{$^{65}$Michigan State University,                                 
            East Lansing, Michigan 48824, USA}                                
\centerline{$^{66}$University of Mississippi,                                 
            University, Mississippi 38677, USA}                               
\centerline{$^{67}$University of Nebraska, Lincoln, Nebraska 68588, USA}      
\centerline{$^{68}$Princeton University, Princeton, New Jersey 08544, USA}    
\centerline{$^{69}$State University of New York, Buffalo, New York 14260, USA}
\centerline{$^{70}$Columbia University, New York, New York 10027, USA}        
\centerline{$^{71}$University of Rochester, Rochester, New York 14627, USA}   
\centerline{$^{72}$State University of New York,                              
            Stony Brook, New York 11794, USA}                                 
\centerline{$^{73}$Brookhaven National Laboratory, Upton, New York 11973, USA}
\centerline{$^{74}$Langston University, Langston, Oklahoma 73050, USA}        
\centerline{$^{75}$University of Oklahoma, Norman, Oklahoma 73019, USA}       
\centerline{$^{76}$Oklahoma State University, Stillwater, Oklahoma 74078, USA}
\centerline{$^{77}$Brown University, Providence, Rhode Island 02912, USA}     
\centerline{$^{78}$University of Texas, Arlington, Texas 76019, USA}          
\centerline{$^{79}$Southern Methodist University, Dallas, Texas 75275, USA}   
\centerline{$^{80}$Rice University, Houston, Texas 77005, USA}                
\centerline{$^{81}$University of Virginia, Charlottesville,                   
            Virginia 22901, USA}                                              
\centerline{$^{82}$University of Washington, Seattle, Washington 98195, USA}  
}                                                                             

\date{May 9, 2007}

\begin{abstract}
We present a study of $ee\gamma$ and $\mu\mu\gamma$ events using 1109 (1009) pb$^{-1}$ of data 
in the electron (muon) channel, respectively. These data were collected with 
the D0~detector at the Fermilab Tevatron $p\bar{p}$ Collider at 
$\sqrt{s}$ = 1.96 TeV. Having observed 453 (515) candidates in the 
$ee\gamma$ ($\mu\mu\gamma$) final state, we measure the $Z\gamma$ 
production cross section for a photon with transverse energy $E_{T} > 7$ GeV, 
separation between the photon and leptons 
$\Delta R_{\ell\gamma}$ $>$ 0.7, and invariant mass of the di-lepton pair 
$M_{\ell\ell}$ $>$ 30 GeV/$c^2$, to be $4.96 \pm 0.30~{\rm (stat. + syst.)} \pm 0.30~{\rm (lumi.)}$ pb, 
in agreement with the standard model prediction of 
$4.74 \pm 0.22$ pb. This is the most precise $Z\gamma$ cross section 
measurement at a hadron collider. We set limits on anomalous trilinear 
$Z\gamma\gamma$ and $ZZ\gamma$ gauge boson couplings of 
$-0.085 < h_{30}^{\gamma} < 0.084$, $-0.0053 < h_{40}^{\gamma} < 0.0054$ 
and  $-0.083 < h_{30}^{Z} < 0.082$, $-0.0053 < h_{40}^{Z} < 0.0054$ at the 
95$\%$ C.L. for the form-factor scale $\Lambda = 1.2$ TeV. 
\end{abstract}

\maketitle

The analysis of vector boson self-interactions provides an important test of 
the gauge sector of the standard model (SM). 
At the tree level, a $Z$ boson cannot couple to a photon.
Various extensions of the SM~\cite{zzg-theory} predict large (anomalous)
values of the trilinear couplings $ZZ\gamma$ and $Z\gamma\gamma$ 
that result in an excess of photons with high transverse energy 
($E_T$) compared with the SM prediction. Consequently, the cross section 
for $Z\gamma$ production will be higher than that predicted by the SM. 
An observation of either an enhancement of the cross section or an
excess of photons with high $E_T$ would indicate new 
physics~\cite{Gounaris_2002za}. 

Previous studies of $Z$ boson production in association with a photon 
were made at the Fermilab Tevatron $p\bar{p}$ Collider by the CDF~\cite{cdf} and 
D0~\cite{d01,prev_zg} collaborations, and at the LEP collider by the 
ALEPH~\cite{aleph}, DELPHI~\cite{delphi}, L3~\cite{L3}, and OPAL~\cite{opal}
collaborations. The combined LEP results are available in 
Ref.~\cite{lep_combined}. All results are consistent with SM predictions. 

In this work, we present a measurement of the $Z\gamma$ cross section 
and a search for anomalous trilinear $ZZ\gamma$ and $Z\gamma\gamma$ couplings.
We follow the framework of Ref.~\cite{baur}, where the $ZV\gamma$ 
($V$ = $Z$, $\gamma$) couplings are assumed only to be 
Lorentz and gauge invariant. Such $ZV\gamma$ couplings can be parameterized by
two CP-violating ($h_1^V$ and $h_2^V$) and two CP-conserving ($h_3^V$
and $h_4^V$) complex parameters. Partial wave unitarity is 
ensured by using a form-factor parameterization that causes the
coupling to vanish at high center-of-mass energy $\sqrt{\hat s}$: 
$h_i^V = \frac{h_{i0}^{V}}{(1+\hat s/\Lambda^{2})^{n}}$.
Here, $\Lambda$ is a form-factor scale, $h^{V}_{i0}$ are the
low-energy approximations of the couplings, and $n$ is the form-factor
power. In accordance with Ref.~\cite{baur}, we set $n = 3$ for 
$h^{V}_{1,3}$ and $n = 4$ for $h^{V}_{2,4}$. In this analysis, we set 
limits only on the real parts of the anomalous couplings, $Re(h_i^V)$. 
For convenience we omit the notation $Re$ and refer to 
symbols $h_i^V$ as the real parts of the couplings throughout~the~text.

The signal sample is selected by requiring a final state that consists
of a photon and a pair of either muon or electron candidates. 
The $e^{+}e^{-}$ and $\mu^{+}\mu^{-}$ pairs can be produced either by
the decay of an on-shell $Z$ boson or via a virtual $Z$ boson or 
a photon through the Drell-Yan production mechanism. We do not 
distinguish between these two processes. The photon can be produced 
by final state radiation (FSR) off both charged leptons or by one of the partons in
the $p$ or $\bar{p}$ through initial state radiation (ISR). We
collectively refer to all these processes as $Z\gamma$ production.

Data for this analysis were collected with the D0~detector
at the Tevatron Collider at $p\bar{p}$ center-of-mass
energy $\sqrt{s}=1.96$ TeV between October 2002 and February 2006.
The integrated luminosity is 1109$\pm$68 (1009$\pm$62)~pb$^{-1}$~\cite{d0lumi} 
for the electron (muon) final state. 

The D0~detector~\cite{run2det} is a multi-purpose detector designed 
to operate at the high luminosity Tevatron collider. The main components 
of the detector are an inner tracker, liquid-argon/uranium
calorimeters, and a muon system. The inner tracker consists of a 
silicon microstrip tracker (SMT) and a central fiber tracker (CFT), 
located in a 2~T superconducting solenoidal magnet and capable of 
providing measurements up to pseudorapidities of $|\eta| \approx$~3.0 
and $|\eta| \approx $~1.8, respectively. The calorimeter is
divided into three sections which cover a wide range 
of pseudorapidities: the central calorimeter (CC) for $|\eta| < $~1.1 
and two end calorimeters (EC) which extend 
coverage to $|\eta|\approx$~4. The calorimeters are longitudinally segmented 
into electromagnetic (EM) and hadronic sections.
The muon system is the outer subsystem of the D0~detector. It includes 
tracking detectors, scintillation trigger counters, and a 1.8~T toroidal magnet, 
and has coverage up to $|\eta| \approx $~2.0. Luminosity is measured 
using plastic scintillator arrays located in front of the EC cryostats 
and covering 2.7~$< |\eta| <$~4.4.

The D0~detector utilizes a three-level (L1, L2, and L3) trigger system. 
In the electron decay channel, we require events to satisfy one of the 
high-$E_T$ electron triggers. At L1, these triggers
require an event to have an energy deposit of more than 10 GeV 
in the EM section of the calorimeter.
At L3, additional requirements are imposed on the fraction
of energy deposited in the EM calorimeter and the shape of
the energy deposition. Single high-$E_T$ triggers are about 99\% efficient
for selecting a pair of electrons from $Z\to ee$ decays.
In the muon decay channel, we require events to satisfy single and di-muon 
triggers. The single muon trigger requires hits in the muon system 
scintillators and a match with a track at L1, and, in portions of the 
data set, also requires spatially-matched 
hits in the muon tracking detectors. At L2, muon track segments are 
reconstructed and a $p_{T}$ requirement is imposed. 
At L3, some of the triggers also require a
reconstructed track in the inner tracker with transverse momentum ($p_T$) 
greater than 10 GeV/$c$. Di-muon trigger requirements on individual
muon candidates are less stringent than those of single muon triggers, 
but they require two muon candidates at L1. 
The muon trigger definitions were changing over the period of time 
when the data set was collected, therefore, to calculate trigger efficiencies 
we divide the data into several subsets and estimate the trigger 
efficiency separately for each subset. 
As determined from the Monte Carlo simulation, the resulting overall 
muon trigger efficiency is 68\% to select a pair of muons
from $Z\to\mu\mu$ decays.

We select $Z$ boson candidates in the electron channel by requiring two 
isolated energy deposits (electromagnetic clusters) corresponding to 
$E_{T} > $~15~GeV in the calorimeter with at least 90\% of their energy
deposited in the EM calorimeter, have a shower shape consistent with that of an 
electron, have matched tracks, and form an invariant mass $M_{ee} > 30$~GeV/$c^2$.
To satisfy single EM trigger requirements, we require at least one of the 
electron candidates to have $E_T >$~25~GeV.
As $Z\gamma$ production yields leptons predominantly at small pseudo-rapidity ($\eta$)
and tracking reconstruction efficiency decreases 
rapidly with $\eta$ in the endcap region, we require at least one
electron candidate to be reconstructed in the central region of
the calorimeter with $|\eta|<1.1$. The other electron can be
reconstructed either in the central region (CC-CC topology) or in
the endcap (CC-EC topology). 

To select $Z\to\mu\mu$ events 
we require the event to have a pair of muon candidates 
each with $p_T$ $>$ 15 GeV/$c$, reconstructed within the muon system 
acceptance, that match to central tracks. At least one 
of the muon candidates must have $p_T$ $>$~20~GeV/$c$. To suppress the background from
hadronic $b\overline{b}$ production, with $b$ quarks decaying
semi-leptonically, we require muon candidates to be isolated from 
other activity in both the central tracker and the calorimeter. 
The background from cosmic ray muons is suppressed by rejecting 
muon tracks that are inconsistent with being produced at the 
interaction point. This background is further reduced by rejecting 
muon candidates that are reconstructed 
back-to-back with an opening angle $\Delta\alpha_{\mu\mu} = 
         |\Delta\phi_{\mu\mu} + \Delta\theta_{\mu\mu} - 2\pi| < 0.05$.    
Both muon candidates must be consistent with being produced at the same 
vertex, i.e., they must originate within 2 cm 
from each other. The event is also rejected if the invariant
mass of the  muon pair $M_{\mu\mu} < 30$~GeV/$c^2$. 

Each event must have at least one photon candidate identified in
the central region of the D0 detector ($|\eta| < 1.1$) 
that deposits at least 90\% of its energy in the EM calorimeter
and has a shower shape consistent with that of a photon.
The photon candidate $E_T$ must exceed 7~GeV, and it must be
separated from both leptons by 0.7 in 
$\Delta R$~=~$\sqrt{(\phi_\ell - \phi_\gamma)^{2} +
  (\eta_{\ell}-\eta_{\gamma})^{2}}$.
As electrons are not a significant source of a background to photon candidates
in $Z\gamma$ final state, we do not require an anti-track match
to the photon candidate. To reduce contamination from EM-like jets, 
we require the photon to be isolated from reconstructed tracks in the annulus of 
0.05~$<$~$\Delta R$~$<$~0.4 -- the scalar sum of all the momenta of the tracks in this
annulus must be below 1.5~GeV/$c$.
These track isolation requirements result in overall improvement (about 8\%)
of photon identification efficiency and smaller systematic uncertainty 
due to simulation of converted photons than that employed
in the previous analysis~\cite{prev_zg}.
The $Z(\gamma)$ candidate events that pass all the
selection criteria are collectively called the signal sample.

We determine the electron and muon identification efficiencies 
using the tag and probe method~\cite{tag_and_probe} on 
$Z \to \ell\ell$ data samples. In the electron channel, we
parameterize the efficiency as a function of $\eta$, $E_{T}$, 
and $z$-coordinate of the interaction vertex.
The reconstruction and trigger efficiencies are then 
calculated using the signal Monte Carlo samples processed 
with the {\sc geant}-based~\cite{geant} D0~simulation package. 
The efficiency to reconstruct a pair of electrons is estimated to 
be (64.6~$\pm$~2.2)\% for the CC-CC topology and 
(50.8~$\pm$~2.4)\% for the CC-EC topology, resulting 
in the combined reconstruction and trigger efficiency 
in the electron channel to be (64.0~$\pm$~2.3)\% for the CC-CC topology 
and (50.3~$\pm$~2.4)\% for the CC-EC topology, respectively. 
The efficiency to reconstruct a pair of muons
is measured to be $(78.8~\pm~1.6)\%$, resulting in (53.4~$\pm$~1.2)\% 
combined reconstruction and trigger efficiency. The main contribution 
to the uncertainty in both channels is from the lepton 
identification uncertainty.
As no clean high-$E_T$ photon sample exists, we measure the photon 
identification efficiency using photon Monte Carlo simulation. 
The quality of how well Monte Carlo simulation describes
EM objects in data is studied on $Z\rightarrow ee$ candidate data sample.
The data and Monte Carlo electron $E_T$ distributions agree at $(99\pm1)\%$.
We normalize the photon efficiency by this number to correct for data-Monte Carlo
simulation difference. The photon identification efficiency is parameterized as
a function of $E_T$; it is measured to rise from $\sim87\%$ 
at a photon $E_T$ of 7~GeV to $\sim95\%$ for photons with $E_T$ 
above 75~GeV.

The main background to the $Z\gamma$ process is $Z$+jet production, 
where an EM-like jet is misidentified as a photon. 
The procedure is to count the number of jets in $Z$+jet events that 
satisfy ``loose'' EM identification criteria and scale that by the
probability for a jet that passes ``loose'' EM requirements to also 
pass the rest of the EM identification criteria. This is slightly 
complicated by the presence of real photons in the jet data. 
We correct for the contribution due to real photons by removing them from 
the sample when determining the jet misidentification rate. 
In detail, to estimate the $Z$+jet background to the $Z\gamma$ process,
we loosen the shower shape and track isolation requirements 
on the photon candidates. If such ``loose'' photon candidates' 
$E_T$ spectrum is denoted as $dN_{\rm EM}/E_T$ and that for 
photon candidates is denoted as $dN_{\gamma}/E_T$, then the 
number of $Z$+jet background events in the signal
sample, $N_{Z+{\rm jet~~bkg}}$, can be determined by using 
the following formula:
\begin{equation}
N_{Z+{\rm jet~~bkg}} = \int\frac{f}{\epsilon_\gamma - f} 
\left(\epsilon_\gamma \frac{dN_{\rm EM}}{E_T} - \frac{dN_\gamma}{E_T}\right)dE_T,
\end{equation} 
where $\epsilon_\gamma$ is the $E_T$-dependent
photon identification efficiency (measured with
respect to the ``loose'' photon identification criteria), 
and $f$ is the $E_T$-dependent 
probability of a jet that satisfies ``loose'' EM criteria 
to pass shower shape requirements.
This probability is determined from a data sample that has 
at least one high-quality jet candidate that satisfies the D0~jet trigger 
requirement. Such data are primarily due to multijet production, and any
photon-like cluster is likely to be a misidentified jet. The
$E_T$-dependent probability is measured as the ratio 
of all EM clusters that pass all photon 
identification criteria to the total number of ``loose''
photon candidates reconstructed in the sample. 
There are real photons in the
sample from direct photon production ($\gamma$+jet processes),
leading to an enhancement in the misidentification
rate for photons with high $E_T$. This contribution is removed
by taking into account the relative cross sections of multijet and
$\gamma$+jet processes and using the $\gamma$+jet Monte Carlo
simulation. The misidentification rate is about 20\% at $E_T$ of
7 GeV and rapidly decreases to about 0.5\% for $E_T > 60$ GeV.
The background suppression was improved, compared with the
previous analysis, by improving the performance of the D0 track
reconstruction software and by increasing track-isolation
requirements imposed on the photon candidate.

We predict $29.5 \pm 4.8(\rm stat.) \pm 3.1(\rm syst.)$ background 
events for the~CC-CC~topology~and $25.7~\pm~3.8(\rm stat.)~\pm~2.5(\rm syst.)$ 
background events for the CC-EC topology in the electron channel. 
Thus, the background in the combined electron channel is estimated to be 
$55.2~\pm~6.1(\rm stat.)~\pm~5.6(\rm syst.)$ events, while that in the muon channel
is $61.3~\pm~6.5(\rm stat.)~\pm~6.2(\rm syst.)$.~Backgrounds~from~
other~processes~are~estimated~to~be~negligible.

We estimate the acceptance and efficiencies of the event selection
criteria using Monte Carlo samples produced with the leading-order 
(LO) $Z\gamma$ generator~\cite{baur} and the simulation of the D0~detector. 
The CTEQ6L1 ~\cite{cteqerror} parton distribution function (PDF) 
set is used. The uncertainty on the acceptance due to the choice 
of the PDF set is estimated to be 4.7\% following the procedures 
described in Ref.~\cite{cteqerror}. We estimate the product of
overall reconstruction efficiency and geometrical acceptance
of selection criteria to be $0.049 \pm 0.003$ for the CC-CC topology 
and $0.026 \pm 0.002$ for the CC-EC topology in the electron channel 
and $0.086 \pm 0.005$~in~the~muon~channel.

After applying all of the event selection criteria, we observe 453 
(308 CC-CC and 145 CC-EC) $ee\gamma$ events, while the 
SM predicts $393.4 \pm 37.6$ (255.7 CC-CC and 137.7 CC-EC) 
signal events with $55.2 \pm 8.3$ 
background events. In the muon channel, we observe 515 events 
compared to an estimated $410.5 \pm 35.9$ SM $\mu\mu\gamma$
events and $61.3 \pm 9.0$ background events. Uncertainty due 
to the PDF choice is the main contributor to the SM signal 
uncertainty. A major contribution to the uncertainty in the 
number of background events is the uncertainty 
in the measurement of the jet misidentification rate.

\begin{figure}
\includegraphics[scale=0.45]{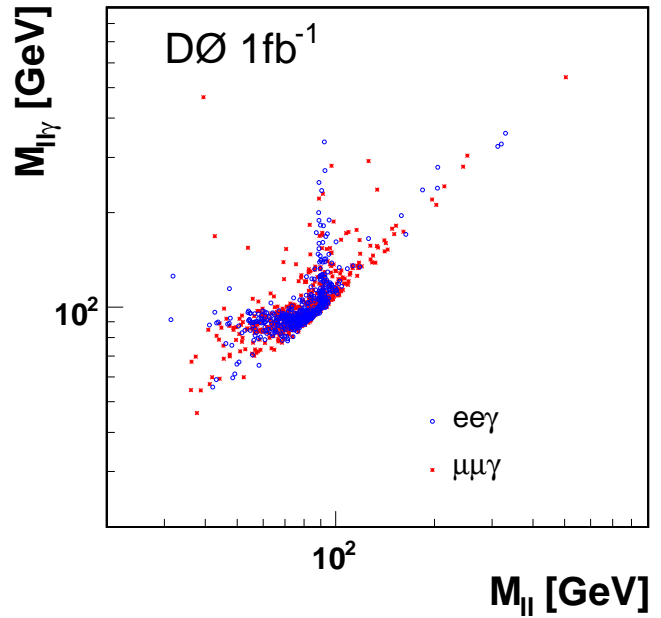}
\caption{Di-lepton+photon $vs.$ di-lepton mass of $Z\gamma$ candidate events.
Masses of candidates in the electron channel are shown as open circles,
while those in the muon mode are shown as stars. 
\label{fig:ll_llg}}
\end{figure}

\begin{figure}
\includegraphics[scale=0.3]{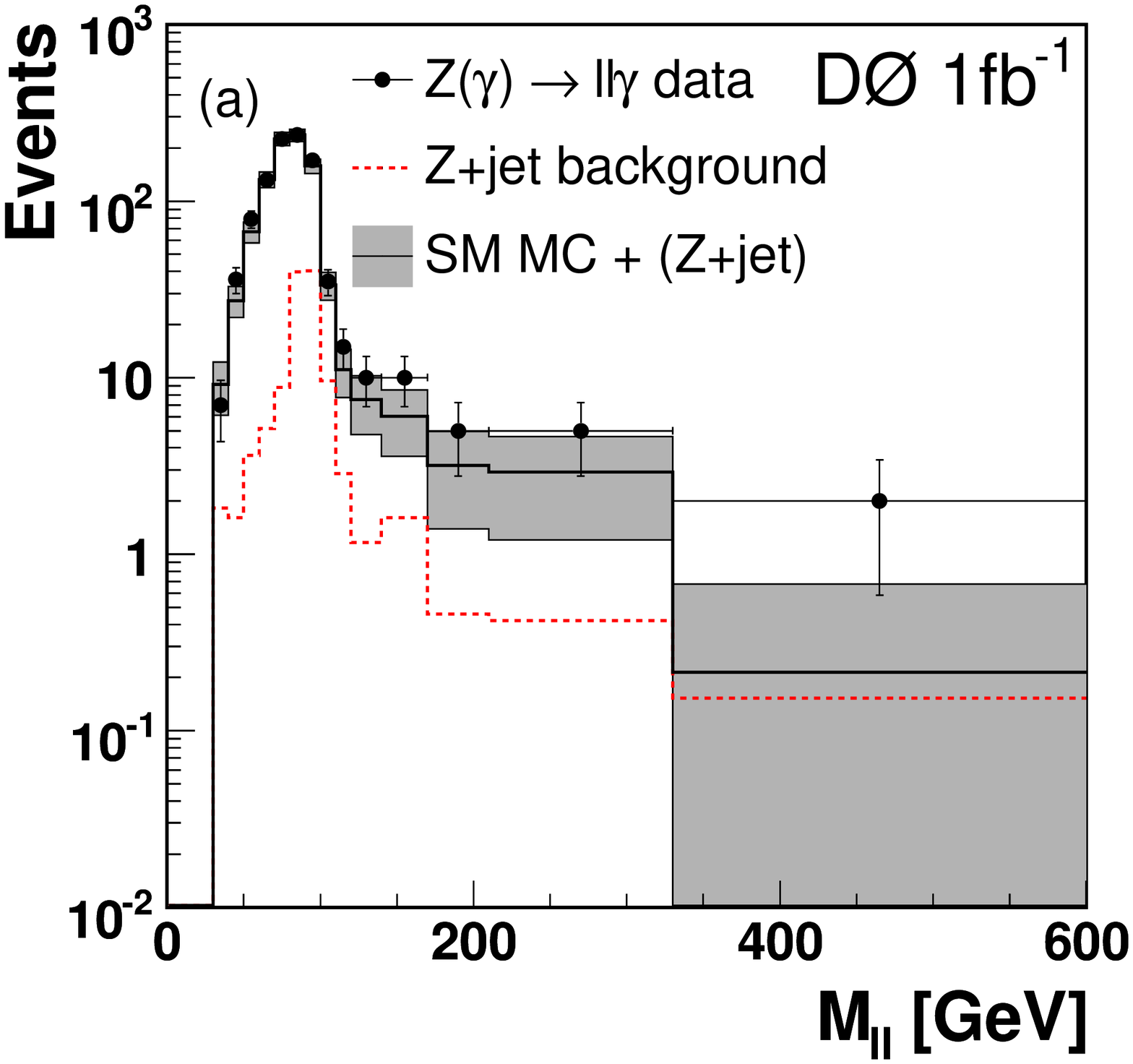}
\includegraphics[scale=0.3]{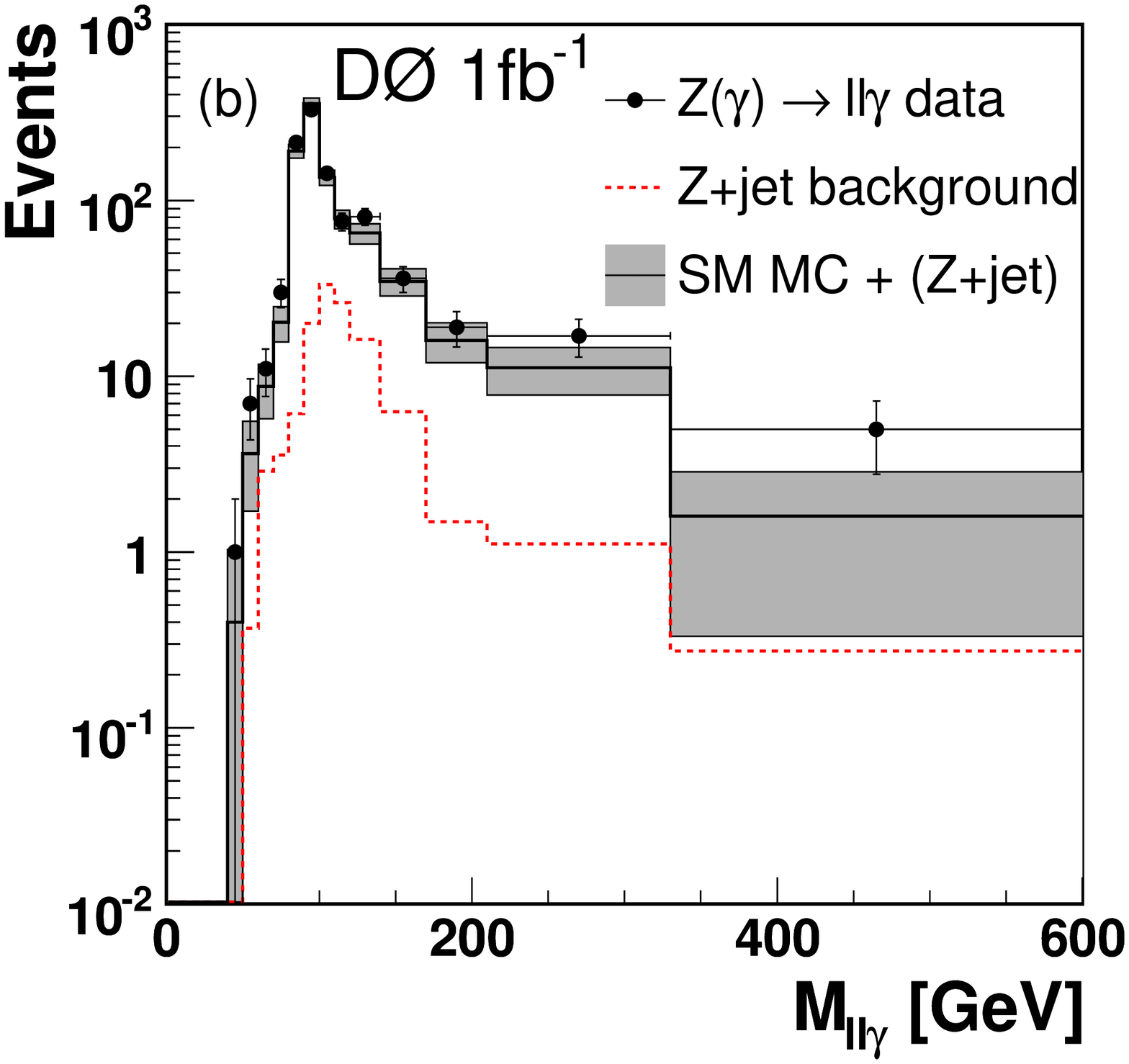}
\caption{(a) Di-lepton mass and (b) di-lepton+photon mass distributions of $\ell\ell\gamma$
data (solid circles), $Z$+jet background (dashed line histogram), and the
standard model plus background (solid line histogram). The shaded bands illustrate 
the systematic and statistical uncertainty on the Monte Carlo and $Z$+jet prediction. 
The Monte Carlo distribution is normalized to the luminosity.
\label{fig:mll_mllg}}
\end{figure}

The invariant mass of di-lepton and photon versus di-lepton invariant mass 
scatter plot is presented in Fig.~\ref{fig:ll_llg}. The  
structure of this distribution reflects the three processes 
through which the final states can be produced. 
Following from the kinematics, the ISR events 
have two leptons from on-shell $Z$ boson decay with 
$M_{\ell\ell}\approx M_Z$ and a photon, emitted by one of the 
interacting partons, resulting in $M_{\ell\ell\gamma}>M_Z$,
and hence the ISR events populate the vertical band. 
The on-shell $Z$ boson FSR events cluster along the horizontal band at 
$M_{\ell\ell\gamma}\approx M_Z$ and have $M_{\ell\ell}<M_Z$. Drell-Yan 
events populate the diagonal band 
with $M_{\ell\ell}\approx M_{\ell\ell\gamma}$.
The di-lepton and three-body mass distributions from data as well as the 
SM prediction with the background overlaid are shown in 
Fig.~\ref{fig:mll_mllg}.

\begin{figure}
\includegraphics[scale=0.4]{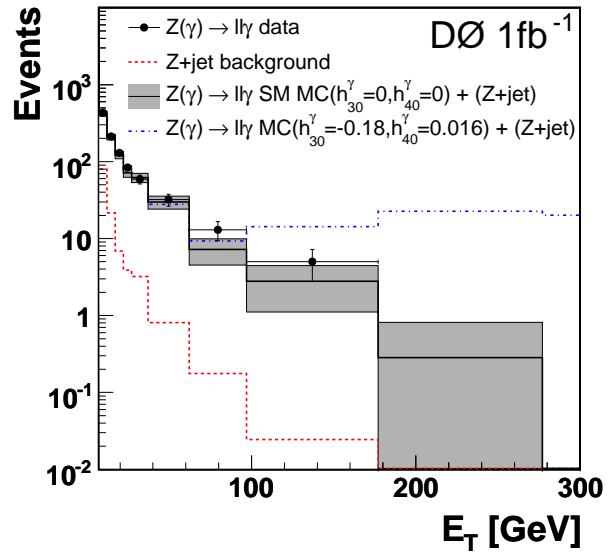}
\caption{
Photon $E_{T}$ spectrum for $\ell\ell\gamma$ data (solid circles), 
$Z$+jet background (dashed line histogram), and Monte Carlo signal plus 
background for the SM prediction (solid line histogram) and for 
the expected distribution when $h^{\gamma}_{30} = -0.18$ and 
$h^{\gamma}_{40} = 0.016$ (dash-dot line histogram). The shaded bands 
illustrate the systematic and statistical uncertainty on the SM Monte Carlo 
and $Z$+jet prediction. The Monte Carlo distributions are normalized 
to the luminosity. 
\label{fig:ptg_all}}
\end{figure}

The measured value of the combined $\ell\ell\gamma$ 
cross section times the branching ratio for $Z(\gamma)\to \ell\ell\gamma$ 
for a photon with $E_{T} > 7$ GeV, separation between the photon and leptons 
$\Delta R_{\ell\gamma}$ $>$ 0.7, and invariant mass of the di-lepton pair 
$M_{\ell\ell}$ $>$ 30 GeV/$c^2$, is
$4.96 \pm 0.30 {\rm (stat.+syst.)} \pm 0.30 {\rm (lumi.)~pb}$. 
The combined statistical and systematic uncertainties contribute to the 
first uncertainty term, and the second uncertainty term is due solely 
to the uncertainty of the luminosity measurement. 
The measured cross section value agrees well with the theoretical 
prediction of $4.74 \pm 0.22$~pb, calculated using the NLO event 
generator~\cite{baurnlo}. 

The $E_{T}$ distribution of the photon candidates in data, compared
with the background and SM prediction is illustrated in
Fig.~\ref{fig:ptg_all}. The $E_T$ distribution expected from 
a new physics process with anomalous couplings is also shown as a 
dashed line. As the measured $Z\gamma$ cross section agrees
well with the SM expectation, we set limits on the real parts of the 
trilinear gauge $ZZ\gamma$ and $Z\gamma\gamma$ couplings 
by comparing the photon candidate $E_T$
distribution, measured in data, with the expected $E_T$ distribution
from anomalous $Z\gamma$ production for a given set of
$ZZ\gamma$ and $Z\gamma\gamma$ coupling values.
The simulation of anomalous $Z\gamma$ production is obtained using the leading-order 
$Z\gamma$ Monte Carlo generator~\cite{baur}. We take into account the
next-to-leading order effects by correcting the leading-order photon
$E_T$ distributions, both for the SM and the anomalous $Z\gamma$ processes, 
with the $E_T$-dependent $K$-factor obtained from the next-to-leading-order 
$Z\gamma$ generator~\cite{baurnlo}.

In this analysis, we set limits on the real parts of CP-conserving anomalous
trilinear couplings $h^{V}_{30}$ and $h^{V}_{40}$ for the form-factor scale 
$\Lambda$~=~1.2~TeV. This choice of $\Lambda$ is not arbitrary, and is 
the highest possible for this current data sample that still ensures 
the limits not to exceed the unitarity boundaries. We generate samples of 
$Z\gamma$ events varying the values of the anomalous couplings
$h_{30}^V$ and $h_{40}^V$, and for each value we compare the photon $E_T$
spectrum from data with that from the simulation with the background
component overlaid.

\begin{figure}
\includegraphics[scale=0.3]{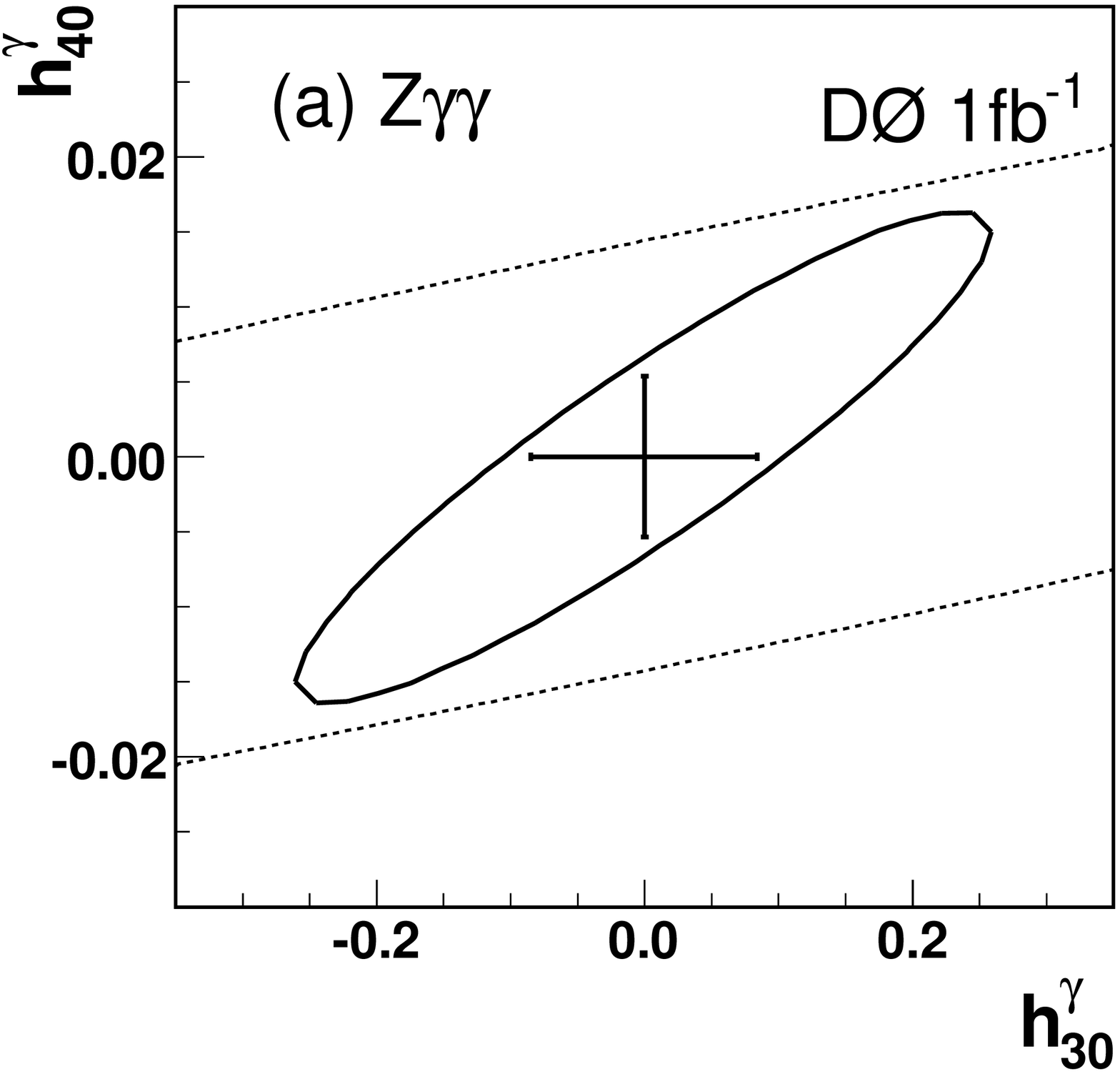}
\includegraphics[scale=0.3]{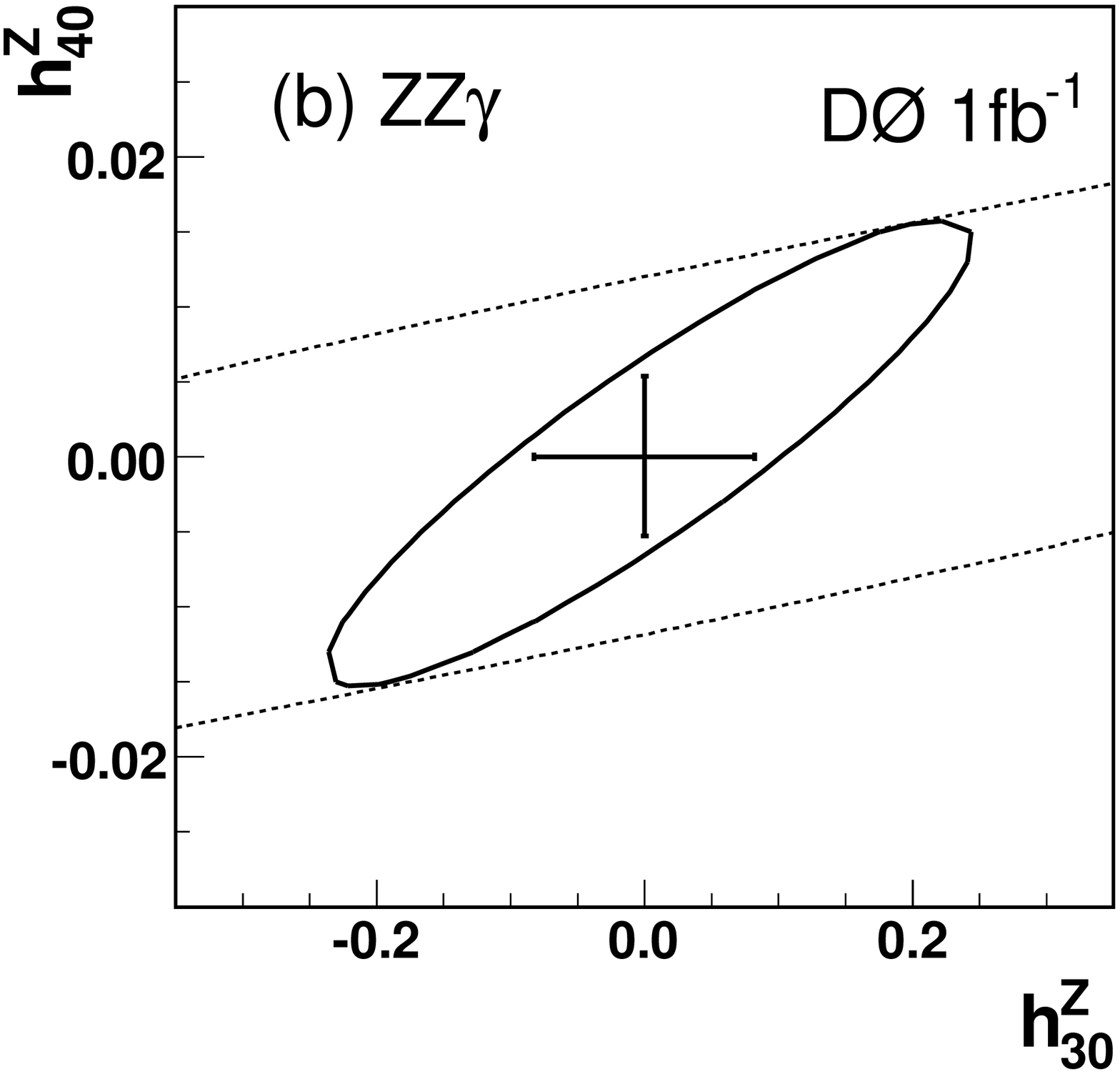}
\caption{The 95\% C.L. two-dimensional contour (ellipse) and one-dimensional 
(ticks on the cross) exclusion limits for the real parts of the CP-conserving
(a) $Z\gamma\gamma$ and (b) $ZZ\gamma$ couplings for $\Lambda$~=~1.2~TeV.
Dashed lines illustrate the unitarity constraints. Both $Z\gamma\gamma$ and $ZZ\gamma$ 
limits are within the unitarity boundaries.
\label{fig:zgglimtev}}
\end{figure}

The likelihood of the data-Monte Carlo simulation match is calculated assuming 
Poisson statistics for the signal (both in the data and MC samples) and the background. 
All systematic uncertainties on backgrounds, efficiencies, and  luminosity are 
taken to be Gaussian. The two-dimensional 95\% C.L. 
limits are shown in Fig.~\ref{fig:zgglimtev}. We also measure 95\% 
C.L. limits on individual anomalous couplings by setting 
the other couplings to their SM value (zero). These limits are presented in 
Table~\ref{tab:LimitSummary} and shown in Fig.~\ref{fig:zgglimtev} with crosses. 
The limit on $h_{10}^V$ ($h_{20}^V$) is the same within the precision of this 
measurement as the limit on $h_{30}^V$ ($h_{40}^V$)~\cite{baurnlo}.
We also obtain one dimensional 95\% 
C.L. limits on the real parts of CP-conserving 
anomalous couplings for the form-factor scale $\Lambda$~=~1~TeV to be
$-0.111 < h_{30}^{\gamma} < 0.113$, 
$-0.0078 < h_{40}^{\gamma} < 0.0079$ and  $-0.109 < h_{30}^{Z} < 0.110$, 
$-0.0077 < h_{40}^{Z} < 0.0078$. This is roughly a factor of two improvement 
compared to the results obtained in Ref.~\cite{prev_zg}. It should be noted 
that  Ref.~\cite{prev_zg} could not use form-factor scale $\Lambda = 1.2$~TeV 
because the resulting anomalous coupling limits would have been outside 
the contours provided by the $S$-matrix unitarity.

\begin{table}
\caption{\label{tab:LimitSummary}Summary of the 95\% C.L. limits on
the real parts of the anomalous couplings for a form-factor scale of $\Lambda =$~1.2~TeV. 
Limits are set by allowing only one coupling to vary; 
the other is fixed to its SM value.}
\begin{ruledtabular}
\begin{tabular}{lcr}
$-0.085 < h_{30}^{\gamma} < 0.084$ & $-0.0053 <  h_{40}^{\gamma} < 0.0054$ & $(h_{i}^{Z} = 0)$\\\\
$-0.083 < h_{30}^{Z} < 0.082$ & $-0.0053 <  h_{40}^{Z} < 0.0054$ & $(h_{i}^{\gamma} = 0)$\\
\end{tabular}
\end{ruledtabular}
\end{table}

In this study we analyzed a sample of 968 $\ell\ell\gamma$ events,
consistent with $Z\gamma$ production. These data correspond
to about 1 fb$^{-1}$ of integrated luminosity, roughly three times more
than what was used in the previous D0~analysis~\cite{prev_zg}. 
This current study also takes advantage of numerous improvements in 
the detector simulation, particle identification, and signal modeling.
The cross section of the $Z\gamma$ process is measured 
to be $4.96 \pm 0.30{\rm (stat. + syst.)} \pm 0.30 {\rm (lumi.)}$~pb. 
This value is consistent with the SM, and is the most 
precise measurement of a $Z\gamma$ cross section at a hadron collider. The observed photon 
$E_T$ distribution, as well as other kinematic parameters, do not indicate 
new physics beyond the SM, allowing us to set limits on the real parts of the 
anomalous $Z\gamma\gamma$ and $ZZ\gamma$ couplings.  
The one dimensional limits at the 95\% C.L. for CP-conserving 
couplings are $-0.085 < h_{30}^{\gamma} < 0.084$, 
$-0.0053 < h_{40}^{\gamma} < 0.0054$ and  $-0.083 < h_{30}^{Z} < 0.082$, 
$-0.0053 < h_{40}^{Z} < 0.0054$ for $\Lambda = 1.2$ TeV. Limits on the CP-violating
couplings are the same as those on the corresponding CP-conserving couplings
within the quoted precision. These new limits represent a significant improvement over 
previous results and the limits on $h^{V}_{40}$ are the most stringent to date.

%
We thank the staffs at Fermilab and collaborating institutions, 
and acknowledge support from the 
DOE and NSF (USA);
CEA and CNRS/IN2P3 (France);
FASI, Rosatom and RFBR (Russia);
CAPES, CNPq, FAPERJ, FAPESP and FUNDUNESP (Brazil);
DAE and DST (India);
Colciencias (Colombia);
CONACyT (Mexico);
KRF and KOSEF (Korea);
CONICET and UBACyT (Argentina);
FOM (The Netherlands);
PPARC (United Kingdom);
MSMT (Czech Republic);
CRC Program, CFI, NSERC and WestGrid Project (Canada);
BMBF and DFG (Germany);
SFI (Ireland);
The Swedish Research Council (Sweden);
Research Corporation;
Alexander von Humboldt Foundation;
and the Marie Curie Program.
%

\end{document}